\documentclass[prc,superscriptaddress,nofootinbib,floatfix]{revtex4}
\usepackage{graphicx}

\usepackage{subfigure}
\newcommand{\beq}{\begin{equation}}
\newcommand{\eeq}{\end{equation}}
\newcommand{\bea}{\begin{eqnarray}}
\newcommand{\eea}{\end{eqnarray}}

\begin{document}





\title{Quantum simulations of strongly coupled quark-gluon plasma 
}
\author{V.S.~Filinov}
\thanks{Corresponding author\quad E-mail:~\textsf{vladimir\_filinov@mail.ru}}
\affiliation{Joint Institute for High Temperatures, Russian Academy of
Sciences, Moscow, Russia}
\author{Yu.B. Ivanov}
\affiliation{GSI Helmholtzzentrum
 f\"ur Schwerionenforschung, 
Darmstadt, Germany}
\affiliation{Kurchatov Institute,
Moscow, Russia}
\author{M. Bonitz}
\affiliation{Institute for Theoretical Physics and Astrophysics, Christian Albrechts University, 
Kiel, Germany}
\author{P.R. Levashov}
\affiliation{Joint Institute for High Temperatures, Russian Academy of
Sciences, Moscow, Russia}
\author{V.E. Fortov}
\affiliation{Joint Institute for High Temperatures, Russian Academy of
Sciences, Moscow, Russia}
%



\begin{abstract}
A strongly coupled quark-gluon plasma (QGP) of heavy constituent quasiparticles
is studied by
 a path-integral Monte-Carlo method, 
which improves the corresponding classical
simulations 
by extending them  to the quantum regime. It is shown that this method is
able to reproduce the lattice equation of state and also
yields valuable insight into the internal structure of the QGP.
The results indicate that the QGP reveals liquid-like rather than gas-like properties.
At temperatures just above the critical one it was found that
bound quark-antiquark states still survive. These states are bound by effective string-like forces.
Quantum effects turned out to be of prime importance in these  simulations.
\end{abstract}

\maketitle

\section{Introduction}\label{s:intro}

 Investigation of properties of the quark-gluon plasma (QGP) is
one of the main challenges of strong-interaction physics, both
theoretically
and experimentally.
Many features of this matter were
experimentally discovered at the Relativistic Heavy Ion Collider
(RHIC) at Brookhaven. The most striking result, obtained from analysis
of these experimental data \cite{shuryak08}, is that the deconfined
quark-gluon matter behaves as an almost perfect fluid rather than a perfect gas,
as it could be expected from the asymptotic freedom.

There are  various approaches to studying QGP.
Each approach has its advantages and  disadvantages.
The most fundamental way to compute
properties of the strongly interacting matter is provided by the lattice QCD \cite{Lattice09,Fodor09,Csikor:2004ik}.
Interpretation of these very complicated computations
requires application of various QCD motivated, albeit schematic, models simulating various aspects of the full theory.
Moreover, such models are needed in cases when the lattice QCD fails, e.g. at large
baryon chemical potentials and out of equilibrium.

A semi-classical approximation, based on a point like quasi-particle picture has been introduced in \cite{LM105}.
It is expected that the main features of non-Abelian
plasmas can be understood in simple semi-classical terms without
the difficulties inherent to a full quantum field theoretical analysis.
Independently the same ideas were implemented in terms of molecular dynamics (MD) \cite{Bleicher99}.
Recently this MD approach was further developed in a series of works
\cite{shuryak1,Zahed}. The MD allowed one to treat soft processes in the QGP which
are not accessible by perturbative means.


A strongly correlated behavior of the QGP is expected to show up in long-ranged spatial correlations of quarks and
gluons which, in fact, may give rise to liquid-like and, possibly, solid-like structures.
This expectation is based on a very similar
behavior observed in electrodynamic plasmas  \cite{shuryak1,thoma04}. This similarity was exploited
to formulate a classical non-relativistic model of a color Coulomb interacting QGP \cite{shuryak1} which was  numerically analyzed by classical MD simulations.
Quantum effects were either neglected or included
phenomenologically via a short-range repulsive correction to the pair potential. Such a rough model may become
a critical issue at high densities.
For temperatures and densities of the QGP considered
in  Ref. \cite{shuryak1} these effects are very important as the quasiparticle
thermal wave length is of order the average interparticle distance.


In this contribution we
extend previous classical nonrelativistic simulations \cite{shuryak1}
based on a color Coulomb interaction to the quantum regime.
We develop an  approach based on path integral Monte Carlo (PIMC)
simulations of the strongly coupled QGP which 
self-consistently takes  into account the Fermi (Bose) statistics of quarks (gluons). 
Following an idea of Kelbg \cite{kelbg}, quantum corrections to the pair potential
are rigorously derived \cite{dusling09}.
This method has been successfully applied to strongly coupled electrodynamic plasmas
\cite{filinov_ppcf01,bonitz_jpa03}. 
Examples are partially-ionized dense hydrogen plasmas, where liquid-like and
crystalline behavior was observed \cite{filinov_jetpl00,bonitz_prl05}. Moreover, also partial ionization effects and pressure ionization
could be studied from first principles \cite{filinov_jpa03}. The same methods have been also applied to electron-hole plasmas in
semiconductors \cite{bonitz_jpa06,filinov_pre07}, including excitonic bound states, which have many similarities to the QGP due to
smaller mass differences as compared to electron-ion plasmas.

First results of applications of the PIMC method to the nonideal QGP
have already been briefly reported in \cite{Filinov:2009pimc}.


\section{Basics of the model}\label{semi:model}

Our model is based on a resummation technique and lattice simulations for 
dressed quarks, antiquarks and gluons interacting via the color Coulomb potential.
The assumptions of the model are similar to those of Ref. \cite{shuryak1}:
%
\begin{description}
 \item[I.] 
 All color quasi-particles are heavy, i.e. their mass ($m$) is higher than the mean kinetic energy
 per particle. For instance, at zero net-baryon density it amounts to $m > T$, where $T$ is a temperature.
 Therefore these particles move non-relativistically. This assumption is based on the analysis of lattice data \cite{Lattice02,LiaoShuryak}. 
 \item[II.] Since the particles are non-relativistic, interparticle interaction is dominated
 by a color-electric Coulomb potential, see Eq. (\ref{Coulomb}).
 Magnetic effects are neglected as sub-leading ones. 
 \item[III.] Relying on the fact that the color representations are large, the color operators 
 are substituted by their average values, i.e. by classical color vectors, the
time evolution of which 
is described
by Wong's dynamics \cite{Wong}.
\end{description}
The quality of these approximations and their limitations were discussed in Ref. \cite{shuryak1}.
Thus, 
this model requires the following quantities as an input:
\begin{description}
\item[1.] the quasiparticle mass, $m$, and
\item[2.] the coupling constant $g^2$.
\end{description}
All the input quantities should be deduced from the lattice data or from an appropriate model simulating these data.


Thus,  we consider a three-component QGP consisting of a number of dressed quarks ($N_q$),
antiquarks $(N_{\bar{q}})$ and   gluons $(N_g)$  represented by quasi-particles.
 In thermal equilibrium
 the average values of these numbers can be found in the grand canonical ensemble defined by  the  temperature-dependent Hamiltonian,
which can be written as ${\hat{H}}={\hat{K}}+{\hat{U}}$. The kinetic and color Coulomb interaction energy of the quasi-particles are
\begin{eqnarray}
\label{Coulomb}
{\hat{K}}=
\sum_i \left[m_i(T,\mu_q)+
\frac{\hat{p}^2_i}{2m_i(T,\mu_q)}\right],
\qquad
{\hat{U}}=\frac{1}{2}\sum_{i,j}
\frac{g^2(|r_i-r_j|,T,\mu_q)
\langle Q_i|Q_j \rangle}{4\pi|r_i-r_j|},
\end{eqnarray}
Here the $Q_i$ denote the Wong's color variable
(8-vector in the $SU(3)$ group), 
$T$ is the temperature and $\mu_q$ is
the quark chemical potential. In fact, the quasi-particle mass and the coupling constant,
as deduced from the lattice data, are functions of $T$ and, in general, $\mu_q$.
Moreover, $g^2$ is a function of distance $r$, which produces a linearly rising potential
at large $r$ \cite{Rich}.

The thermodynamic properties in the grand canonical ensemble with given temperature $T$,
chemical potential $\mu_q$ and fixed volume $V$ are fully described by the
grand partition function
\begin{eqnarray}\label{Gq-def}
&&Z\left(\mu_q,\beta,V\right)=
\sum_{N_q,N_{\bar{q}},N_g}\frac{\exp(\mu_q(N_q-N_{\bar{q}})/T)}{N_q!N_{ \bar{q}}!N_g!} \sum_{\sigma}\int\limits_V
dr dQ \,\rho(r,Q, \sigma ; N_q,N_{\bar{q}},N_g;\beta),
\end{eqnarray}
%
%
%
where $\rho(r,Q, \sigma ; N_q,N_{\bar{q}},N_g;\beta)$ denotes the diagonal matrix
elements of the density operator ${\hat \rho} = \exp (- \beta{\hat H})$, and $\beta=1/T$.
Here $\sigma$, $r$ and $Q$
denote the spin, spatial and color degrees of freedom 
of all quarks, antiquarks and gluons in the ensemble, respectively.
Correspondingly, the $\sigma$ summation and integration $dr dQ$ run over
all individual degrees of freedom of the particles.
%
Since the masses and the coupling constant depend on the temperature and chemical potential,
special care should be taken to preserve thermodynamical consistency of this approach.
In order to preserve the thermodynamical consistency,
thermodynamic functions such as pressure, $P$, entropy, $S$, baryon number, $N_B$, and
internal energy, $E$, should be calculated through respective derivatives of
 the logarithm of the partition function
\begin{eqnarray}
\label{p_gen}
P=\partial (T\ln Z) / \partial V, \quad
S=\partial (T\ln Z) / \partial T, \quad
N_B=(1/3)\partial (T\ln Z) / \partial \mu_q, \quad
E= -PV+TS+3 \mu_q N_B.
\end{eqnarray}
This is a conventional way of maintaining the thermodynamical consistency in approaches
of the Ginzburg-Landau type as they are applied in high-energy physics.

The exact density matrix of interacting quantum
systems can be constructed using a path integral
approach~\cite{feynm,zamalin}
based on the operator identity
$e^{-\beta {\hat H}}= e^{-\Delta \beta {\hat H}}\cdot
e^{-\Delta \beta {\hat H}}\dots  e^{-\Delta \beta {\hat H}}$,
where the r.h.s. contains $n+1$ identical factors with $\Delta \beta = \beta/(n+1)$.
which allows us to
rewrite\footnote{For the sake of notation convenience, we ascribe superscript $^{(0)}$
to the original variables.}
the integral in Eq.~(\ref{Gq-def})
\begin{eqnarray}
&&
\sum_{\sigma} \int\limits dr^{(0)}dQ^{(0)}\,
\rho(q^{(0)},Q^{(0)},\sigma; N_q,N_{\bar{q}},N_g;\beta) =
\int\limits  dr^{(0)}dQ^{(0)} dr^{(1)}dQ^{(1)}\dots
dr^{(n)}dQ^{(n)} \, \rho^{(1)}\cdot\rho^{(2)} \, \dots \rho^{(n)}
\nonumber\\&\times&
\sum_{\sigma}\sum_{P_q} \sum_{P_{ \bar{q}}}\sum_{P_g}(- 1)^{\kappa_{P_q}+ \kappa_{P_{\bar{q}}}} \,
{\cal S}(\sigma, {\hat P_q}{\hat P_{ \bar{q}}}{\hat P_g} \sigma^\prime)\,
{\hat P_q} {\hat P_{ \bar{q}}}{\hat P_g}\rho^{(n+1)}\big|_{r^{(n+1)}= r^{(0)}, \sigma'=\sigma} \,
\nonumber\\&=&
\int\limits dQ^{(0)}dr^{(0)} dr^{(1)}\dots dr^{(n)}
\tilde{\rho}(r^{(0)},r^{(1)}, \dots r^{(n)};Q^{(0)}; N_q,N_{\bar{q}},N_g;\beta).
 \label{Grho-pimc}
\end{eqnarray}
The spin gives rise to the spin part of the density matrix (${\cal
S}$) with exchange effects accounted for by the permutation
operators  $\hat P_q$, $\hat P_{ \bar{q}}$ and $\hat P_g$ acting on the quark, antiquark and gluon spatial $r^{(n+1)}$
and color $Q^{(n+1)}$ coordinates,
as well as on the spin projections $\sigma'$. The
sum runs over all permutations with parity $\kappa_{P_q}$ and
$\kappa_{P_{ \bar{q}}}$. In Eq.~(\ref{Grho-pimc}) the index $l=1\dots n+1$
labels the off-diagonal
density matrices
$\rho^{(l)}\equiv \rho\left(r^{(l-1)},Q^{(l-1)};r^{(l)},Q^{(l)};\Delta\beta\right) \approx
\langle r^{(l-1)}|e^{-\Delta \beta {\hat H}}|r^{(l)}\rangle\delta_\epsilon(Q^{(l-1)}-Q^{(l)})$, where
$\delta_\epsilon(Q^{(l-1)}-Q^{(l)})$ is a delta-function at $\epsilon\rightarrow 0$.
Accordingly each $a$ particle is represented by a set of $n+1$ coordinates
(``beads''),
i.e. by $(n+1)$ 3-dimensional vectors
$\{r_a^{(0)}, \dots r_a^{(n)}\}$
and a 8-dimensional color
vector $Q^{(0)}$ in the $SU(3)$ group, since all beads
are characterized by the same color charge.

The main advantage of decomposition (\ref{Grho-pimc}) is that it
allows us to use a  semi-classical  approximation for density matrices $\rho^{(l)}$,
which is applicable due to smallness of artificially introduced factor $1/(n+1)$.
This parameter makes the thermal  wavelength $\Delta\lambda_a=\sqrt{2 \pi \Delta\beta/m_a}$
of a bead of type $a$ ($a = q, \overline{q}, g$),
smaller then a characteristic scale
of variation of the potential energy.
In the high-temperature limit $\rho^{l}$ can be approximated by a product of
two-particle density matrices.
Generalizing the
electrodynamic plasma results \cite{filinov_ppcf01} to the case of an additional bosonic species (i.e. gluons),
we write
\begin{eqnarray}
&&
\tilde{\rho}(r^{(0)},r^{(1)}, \dots r^{(n)};Q^{(0)}; N_q,N_{\bar{q}},N_g;\beta)
\nonumber\\&=&
 \sum_{s,k}\frac{C^s_{N_q}}{2^{N_q}} \frac{C^k_{N_{ \bar{q}}}}{2^{N_{\bar{q}}}}
\frac{\exp\{-\beta U(r,Q,\beta)\}}{\lambda_q^{3N_q} \lambda_{{ \bar{q}}}^{3N_{ \bar{q}}}\lambda_g^{3N_g}}\,
 \, {\rm per}\,||\tilde{\phi}^{n,0}||_{\rm glue} \,
{\rm det}\,||\tilde{\phi}^{n,0}||_s \, {\rm det}\,||\tilde{\phi}^{n,0}||_k \,
\prod\limits_{l=1}^n \prod\limits_{p=1}^N
\phi^l_{pp} \, \label{Grho_s}
\end{eqnarray}
where $N=N_q+N_{\bar{q}}+N_g$, $s$ and $k$ are numbers of quarks  and antiquarks, respectively,
with the same spin projection, $\lambda_a=\sqrt{2 \pi \beta / m_a}$
 is a thermal  wavelength of an $a$ particle,
$C^s_{N_a}=N_a!/[s!(N_a-s)!]$, the antisymmetrization and
symmetrization are taken into account by the symbols ``det'' and ``per'' denoting the determinant and permanent, respectively.
Functions
$\phi^l_{pp}\equiv \exp\left[-\pi\left|\xi^{(l)}_p\right|^2\right]$ and matrix elements
$\tilde{\phi}_{to}^{n,0}=\exp \left(-\pi
\left|(r_{t}^{(0)}-r_{o}^{(0)})+ y_{t}^{(n)}\right|^2/\Delta\lambda_{a}^2\right)
\delta_\epsilon(Q_t-Q_o)$, where $t$ and $o$ are particle's indexes, are expressed in
terms of
distances ($y_{a}^{(1)}, \dots , y_{a}^{(n)}$) and
dimensionless distances ($\xi_{a}^{(1)}, \dots , \xi_{a}^{(n)}$) between
neighboring beads of an $a$ particle, defined as
$r_{a}^{(l)} = r_{a}^{(0)}+y_{a}^{(l)}$, ($l>0$),
and $y_a^{(l)}=\Delta\lambda_a\sum_{k=1}^{l}\xi_a^{(k)}$.
Notice that the $||\tilde{\phi}^{n,0}||$ matrix consists of
three nonzero blocks related to quarks, $||\tilde{\phi}^{n,0}||_s$,
antiquarks, $||\tilde{\phi}^{n,0}||_k$, and gluons, $||\tilde{\phi}^{n,0}||_{\rm glue}$.
%
The density matrix (\ref{Grho_s}) has been transformed to a form which does not
contain an explicit sum over permutations.  
Let us stress that the determinants depend also on the color variables. 

In Eq.~(\ref{Grho_s}) the total color interaction energy
\begin{eqnarray}
U(r,Q,\beta) =
\frac{1}{2(n+1)}\sum_{p\neq t}\sum_{l=1}^{n+1}
\Phi^{pt}(|r_p^{(l-1)}-r_t^{(l-1)}|,|r_p^{(l)}-r_t^{(l)}|, Q_p,Q_t)
\label{up}
\end{eqnarray}
is defined in terms of
off-diagonal two-particle effective quantum potential
$\Phi^{pt}$,
which is obtained by expanding the two-particle density matrix $\rho_{pt}$
up to the first order in small parameter $1/(n+1)$:
\begin{eqnarray}
&&
\rho_{pt}(r,r',Q_p,Q_t,\Delta\beta) \approx \rho_{pt}^0(r,r',Q_p,Q_t,\Delta\beta)-
\int_0^1 d\tau \int dr'' 
\frac{\Delta\beta g^2(|r''|,T,\mu_q)\langle Q_p|Q_t \rangle}{4\pi|r''|\Delta\lambda_{pt}^2\sqrt{\tau(1-\tau)}}
\,\nonumber\\&\times&
\exp\left(-\frac{\pi|r'-r''|^2}{\Delta\lambda_{pt}^2(1-\tau)}\right)
\exp\left(-\frac{\pi|r''-r|^2}{\Delta\lambda_{pt}^2\tau}\right)
\approx \rho_{pt}^0
\exp[-\Delta\beta
\Phi^{pt}(r,r', Q_p,Q_t)].
\label{GPERT}
\end{eqnarray}
where $ r= r_p-r_t$, $r'= r'_p-r_t'$, $\Delta\lambda_{pt}=\sqrt{2\pi\Delta\beta /m_{pt}}$,
$m_{pt}=m_{p}m_{t}/(m_{p}+m_{t})$, and $\rho_{pt}^0$ is the  two-particle
density matrix of the ideal gas.
The result for the diagonal color Kelbg potential can be written as
\begin{eqnarray}
\Phi^{pt}( r,r,Q_p,Q_t) 
\approx \frac{g^2(T,\mu_q)\,\langle Q_p|Q_t \rangle}{4 \pi \Delta\lambda_{pt} x_{pt}} \,\left\{1-e^{-x_{pt}^2} +
\sqrt{\pi} x_{pt} \left[1-{\rm erf}(x_{pt})\right] \right\},
\label{kelbg-d}
\end{eqnarray}
where $x_{pt}=| r_{p}-r_{t}|/\Delta\lambda_{pt}$.
Here the function $g^2(T, \mu_q) = \overline{g^2(r'',T. \mu_q)}$,
resulting from averaging of the initial
$g^2(r'', T, \mu_q)$ over relevant distances of order $\Delta\lambda_{pt}$,
plays the role of an effective coupling constant.
Note that the color Kelbg
potential approaches the color Coulomb potential
at distances larger than $\Delta\lambda_{pt}$. What is of prime importance, the color Kelbg
potential is finite at zero distance, thus removing
in a natural way the classical divergences and making any artificial cut-offs obsolete.
This potential 
is straightforward generalizations of the corresponding potentials of electrodynamic plasmas
\cite{afilinov_pre04}. 
The off-diagonal 
elements of the effective interaction are approximated
 by the diagonal one by means of
$\Phi^{pt}(r,r';,Q_p,Q_t)\approx [\Phi^{pt}(r,r,Q_p,Q_t) + \Phi^{pt}(r',r',Q_p,Q_t)]/2$.

The described path-integral representation of the density matrix
is exact in the limit $n\to \infty$. For any finite
number $n$, the error of the above approximations for the whole product on the r.h.s. of Eq.
(\ref{Grho-pimc}) is of the order $1/(n+1)$ whereas the error of each $\rho^{l}$ is
of the order $1/(n+1)^2$, as it was shown in Ref. \cite{filinov_ppcf01}.

The main contribution to the partition function comes from
configurations in which the `size' of the cloud of beads of quasiparticles is of
the order of their thermal  wavelength
$\lambda_a$
whereas characteristic
distances  between beads of each quasiparticle are of the order of
$\Delta\lambda_a$.

\section{Testing the method within the canonical ensemble}\label{s:model}

To test the developed approach we consider the QGP 
only at zero baryon density 
and further
simplify the model by additional approximations, similarly to Ref. \cite{shuryak1}:
\begin{description}
\item[IV] We replace the grand canonical ensemble by a canonical one.
The thermodynamic properties in the canonical ensemble with given temperature $T$ and fixed volume $V$ are fully
described by the density operator ${\hat \rho} = e^{-\beta {\hat H}}$ with the partition function
defined as follows
\begin{equation}\label{q-def}
Z(N_q,N_{ \bar{q}},N_g,V;\beta) = \frac{1}{N_q!N_{ \bar{q}}!N_g!} \sum_{\sigma}\int\limits_V
dr dQ\,\rho(r,Q, \sigma ;\beta),
\end{equation}
with $N_q=N_{ \bar{q}}$ and hence $N_B=0$.
In order to preserve the thermodynamical consistency of this formulation,
thermodynamic quantities
should be calculated through respective derivatives of
the logarithm of the partition function similarly to that in Eq. (\ref{p_gen})
with the exception that now $N_a$ are indepenent variables.

\item[V] Since the masses of quarks of different flavors extracted from lattice data are very similar,
we do not distinguish between quark flavors.
Moreover, we take the quark and gluon quasi-particle masses being equal because their values
deduced from the lattice data \cite{Lattice02,LiaoShuryak} are very close.
\item[VI]  Because of the equality of masses and approximate equality of number of degrees of freedom
of quarks, antiquarks and gluons, we assume that these species
are equally
represented in the system: $N_q = N_{\bar{q}} = N_g$.
\item[VII] For the sake of technical simplicity, the SU(3) color group is replaced by SU(2).
\end{description}
Thus, this simplified model requires an additional quantity as an input:
\begin{description}
\item[3.] the density of quasi-particles $(N_q+N_{\bar{q}}+N_g)/V=n(T)$ as a function of the temperature.
\end{description}
Although this density is unknown from the QCD lattice calculations and  we use it as a fit
parameter, it is very important to partially overcome constrains of the above simplifications.
First, it concerns the use of the SU(2) color group, which first of all reduces the degeneracy factors of the
quark and gluon states, as compared to the SU(3) case, and thereby reduces pressure and all other thermodynamic
quantities. A proper fit of the density allows us to remedy this deficiency of the normalization.
Second, in fact we consider the system of single quark flavor, i.e. all quarks are identical,
which also reduces the normalization of all thermodynamic quantities. The density fit cures
the deficiency of this normalization, though the excessive anticorrelation of quarks remains.



\begin{figure}[htb]\label{fig:cor}
\vspace{0cm} \hspace{0.0cm}
\includegraphics[width=5.3cm,clip=true]{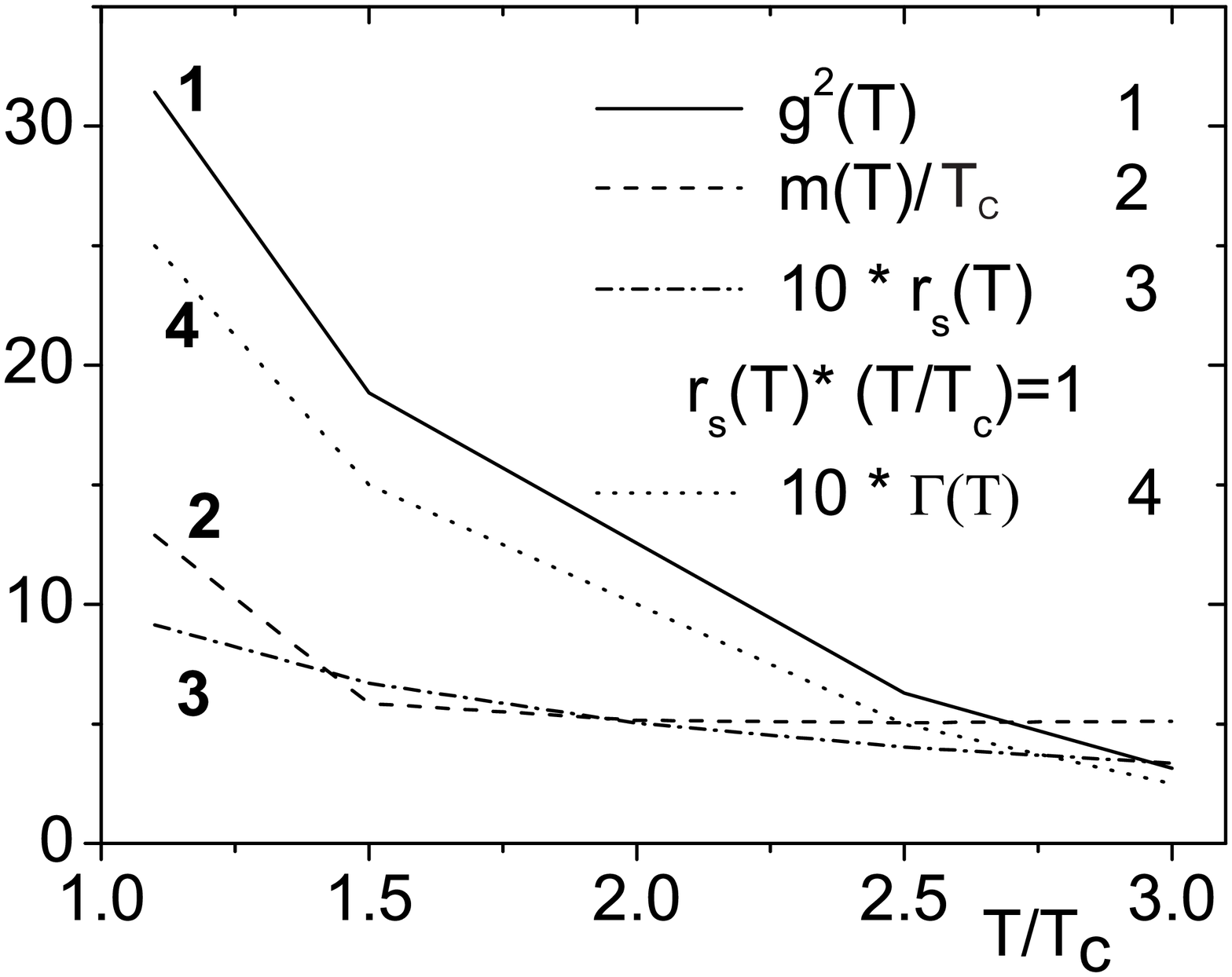}
\includegraphics[width=7.cm,clip=true]{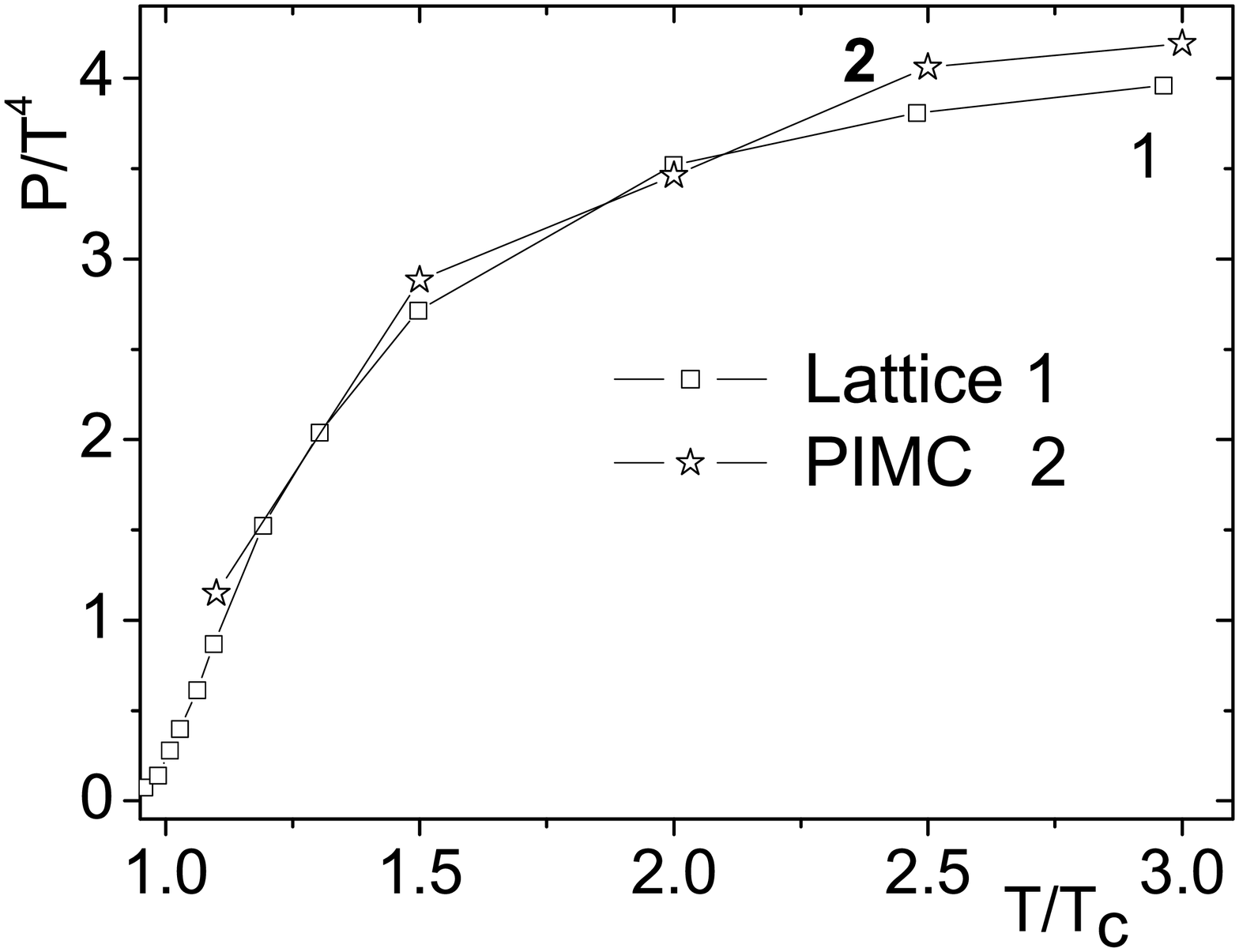}
\caption{Left panel: Temperature dependence of the model input quantities and
the plasma coupling parameter $\Gamma$. 
%
Right panel:
Equation of state (pressure versus temperature) of the QGP from PIMC simulations compared
to lattice data of Refs.~\cite{Lattice09,Csikor:2004ik}.
}
\label{fig:EOS}
\end {figure}

Ideally the parameters of the model should be deduced from the QCD lattice data. However, presently
this task is still quite ambiguous. Therefore, in the present simulations we take a possible (maybe,
not the most reliable) set of parameters.
Following Refs. \cite{LiaoShuryak,shuryak1}, the parametrization
of the quasi-particle mass is taken in the form
\begin{eqnarray}
\label{mass}
m(T)/T_c=0.9/(T/T_c-1)+3.45+0.4T/T_c
\end{eqnarray}
where $T_c=175$ MeV is the critical temperature. This parametrization fits the quark mass at two values
of temperature obtained in the lattice calculations \cite{Lattice02}. According to \cite{Lattice02} the masses
are quite large: $m_q/T \simeq 4$  and $m_g/T \simeq 3.5$. These are essentially larger than masses
required for quasi-particle fits \cite{Peshier96,Ivanov05} of the lattice  thermodynamic properties
of the QGP:  $m_q/T \simeq 1\div 2$  and $m_g/T \simeq 1.5\div 3$.
Moreover, the pole quark mass $m_q/T \simeq 0.8$ was reported
in recent work \cite{Karsch09a}, as deduced from lattice calculations. Nevertheless, in spite of the fact
that it obviously produces too high masses, we use parametrization
(\ref{mass}) in order to be compatible with the input of classical MD
of Ref.  \cite{shuryak1}. 
The $T$-dependence of this mass is illustrated in Fig.~\ref{fig:EOS} (left panel).

The coupling constant used in the simulations is displayed in Fig.~\ref{fig:EOS} as well.
From the point of view of
the QCD phenomenology \cite{Prosperi} it is too high at low energies, i.e. at $T/T_c \simeq 1\div 2$.
However, the high values of masses require such a large value of $g^2$, e.g., to be consistent with the
HTL results for the quasi-particle masses.
The large value of $g^2$ is also required to simulate larger values of Casimirs
(defining the normalization of the color vectors)
in the $SU(3)$ group as
compared to the $SU(2)$ one used here.
Moreover, such high $g^2$ are not inconsistent with the lattice data
\cite{Karsch04}.

The density of quasi-particles, which is additionally required within the canonical-ensemble
approach, was chosen on the condition of the best agreement of the calculated pressure with the corresponding
lattice result, see Fig.~\ref{fig:EOS} (right panel). It was taken to be $n(T)=0.24 T^3$. From the first
glance, it is a very low density. For example, in the classical simulations of Ref. \cite{shuryak1} it was taken
as $n(T)/T^3=6.3$, which corresponds to the density of an ideal gas of {\em massless} quarks, antiquarks
and gluons. Since the quasi-particles are very heavy in the present model (as well as in that of
Ref. \cite{shuryak1}), 
the latter density looks unrealistically high. Even in
quasi-particle models \cite{Peshier96,Ivanov05}, where the masses are lower,
the density turns out to be $n(T)/T^3 \approx 1.4$.
Since Eq. (\ref{mass}) gives even
larger masses than those in Refs. \cite{Peshier96,Ivanov05} and in view of the adopted large coupling,
the chosen value of $n(T)$ does not
look too unrealistic. The $T$-dependence of $n(T)$ is presented in Fig.~\ref{fig:EOS} (left panel)
in terms of the mean interparticle distance $r_s (T)$ in units of $\sigma=1/T_c=1.1$ fm (Wigner-Seitz radius).
Notice that the present choice of $n(T)$ corresponds to the relation $T r_s (T)=1$.

Thus, although the chosen set of parameters is still debatable, it is somehow self-consistent.
In the future we are going to get rid of the $n(T)$ parameter, by applying the grand-canonical approach,
and by using more moderate (and maybe realistic) sets of parameters.

Calculation of the equation of state (right panel of Fig.~\ref{fig:EOS})
was used to optimize the parameters of the model in order to
proceed to predictions of other properties concerning the internal structure
and in the future also non-equilibrium dynamics of the QGP.
The plasma coupling parameter, 
$\Gamma$ defined as a ratio of the average potential to average kinetic energy,
is also presented in
Fig.~\ref{fig:EOS} (left panel). It turns out to be of the order of unity which indicates
that the QGP is a strongly coupled Coulomb liquid rather than a gas.
In the studied temperature range, $1<T/T_c<3$, the QGP  is, in fact, quantum degenerate, since 
the degeneracy parameter
$\chi_a = n_a\lambda_a^3$
varies from $0.1$ to $2$. 

%


%
\begin{figure}[htb]
\vspace{0cm} \hspace{0.0cm}
\includegraphics[width=7.cm,clip=true]{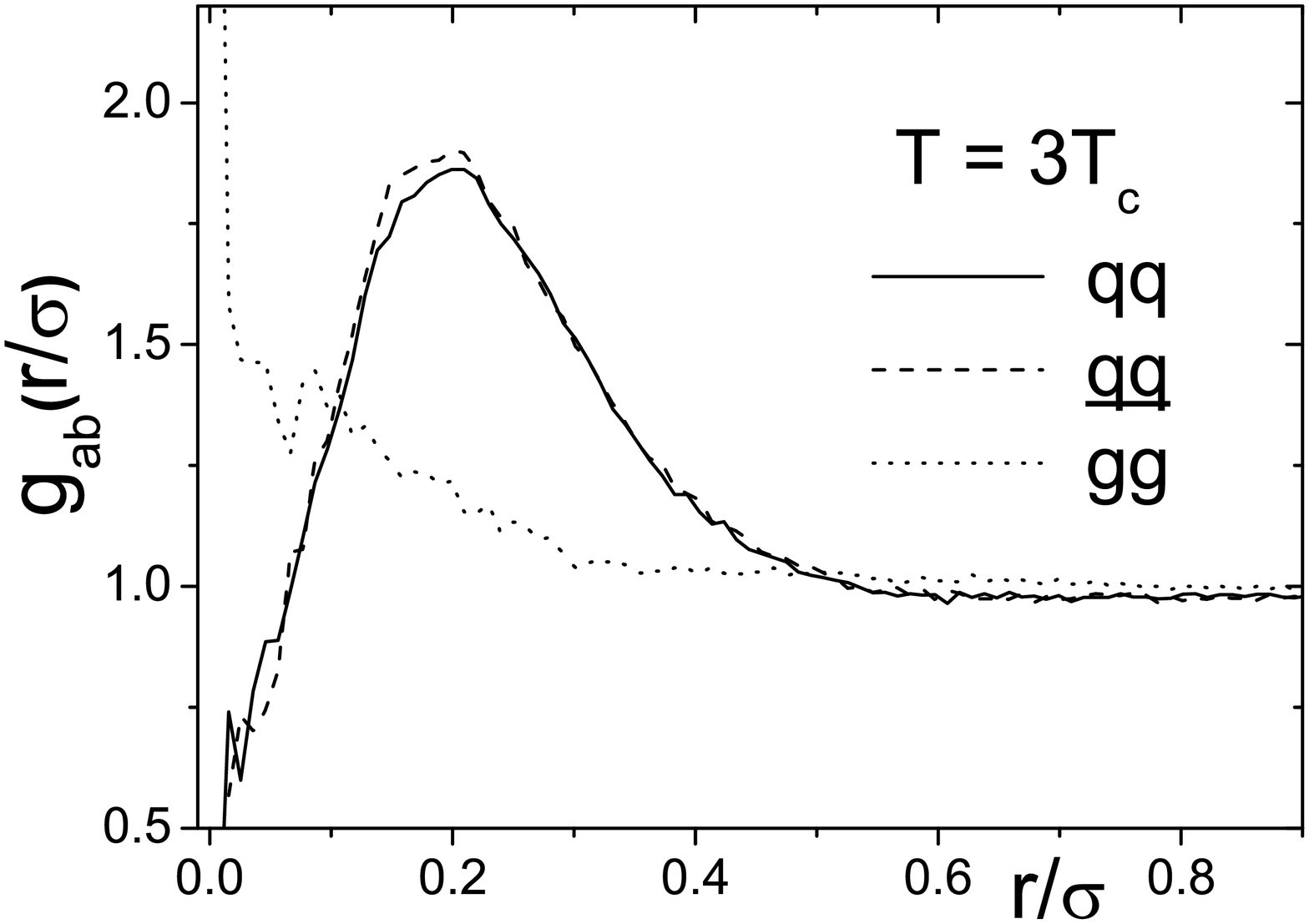}
\includegraphics[width=7.cm,clip=true]{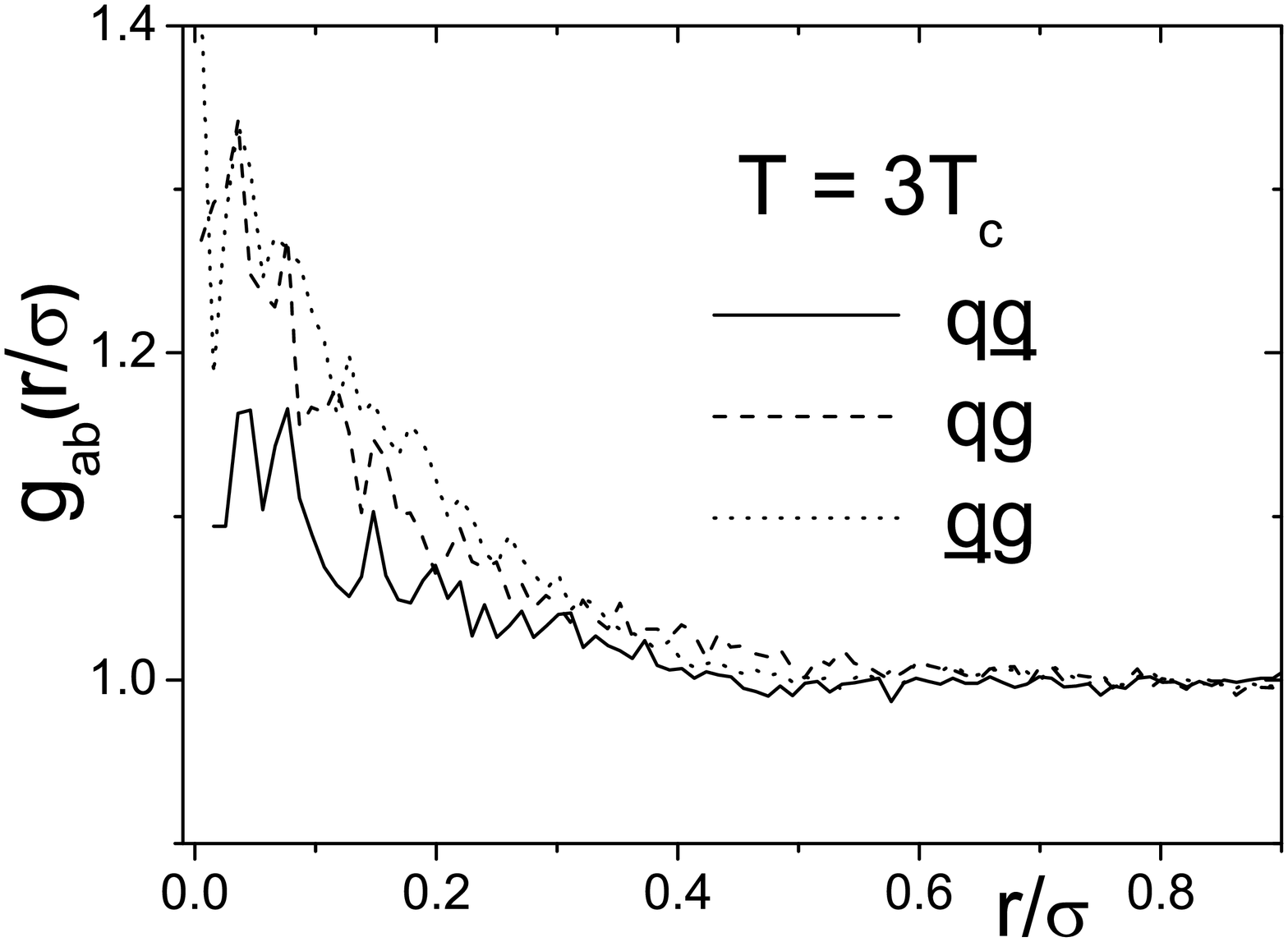}
\includegraphics[width=7.cm,clip=true]{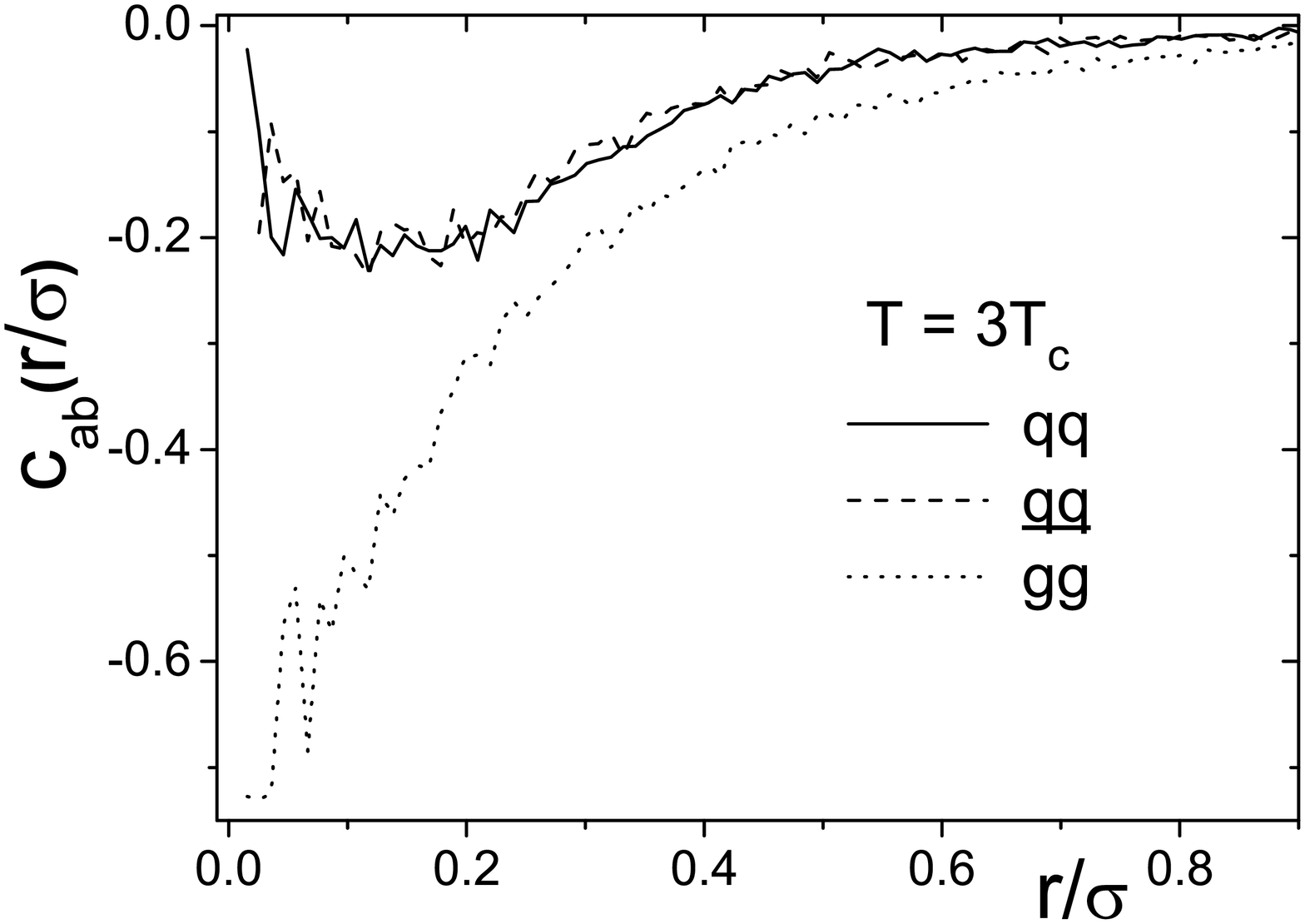}
\includegraphics[width=7.cm,clip=true]{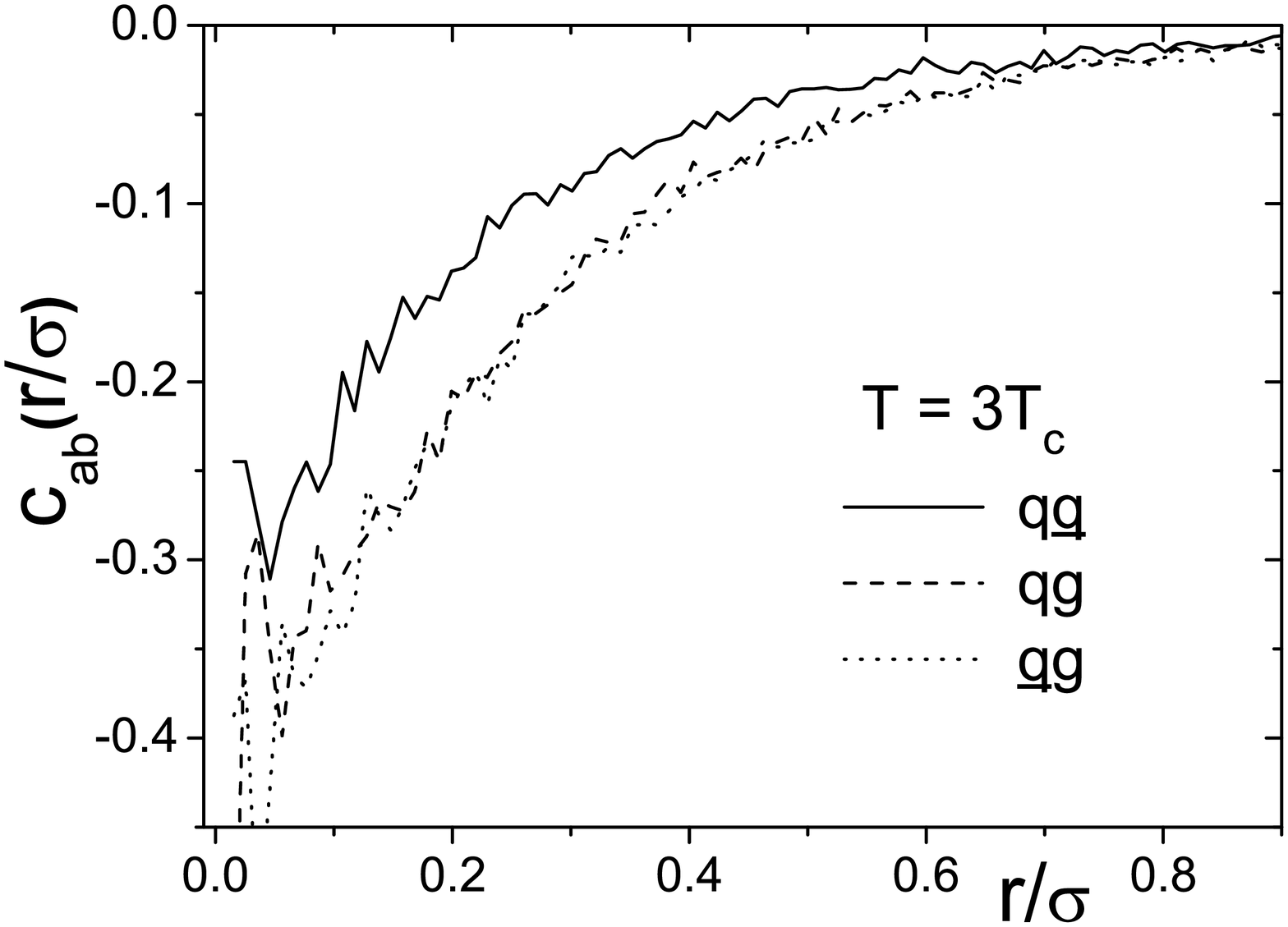}
\caption{Pair distribution functions (upper panels) and color pair distribution
functions (lower panels) of identical (left panels) and different (right panels) quasiparticles  at
temperature $T/T_c=3$,
$\sigma = 1/T_c = 1.1$ fm.
}
\label{fig:PDFC}
\end{figure}
Let us now consider the spatial arrangement of the quasiparticles in the QGP by studying the
 pair distribution functions (PDF's) $g_{ab}(r)$. They give the probability density
to find a pair of particles of types $a$ and $b$ at a certain distance $r$ from each other and are
 defined as
\begin{eqnarray}\label{g-def}
g_{ab}(R_1-R_2)=
\frac{1}{Z N_q!N_{ \bar{q}}!N_g!}
\sum_{\sigma}\int
dr dQ\,\delta(R_1-r^a_1)\delta(R_2-r^b_2)\rho(r,Q, \sigma ;\beta).
\end{eqnarray}
The PDF's depend only on the difference of cordinates because of the translational invariance of the system.
In a non-interacting classical system,
$g_{ab}(r)\equiv 1$, whereas interactions and  quantum statistics result in
a re-distribution of the particles.
Results for the PDF's at temperature $T/T_c=3$ are shown in 
the top panels of Fig.~\ref{fig:PDFC}.
%

At large distances, $r/\sigma \ge 0.5$,
all PDF's of identical particles (top left panel of Fig. \ref{fig:PDFC})
coincide, approaching unity. 
At small distances,
the gluon PDF increases
monotonically when the distance goes to zero, while 
the PDF of quarks (and antiquarks) exhibits a broad maximum.
In the present conditions, the thermal wavelength $\lambda$  approximately equals
$0.37\sigma$, i.e. the difference starts to appear at distances of the order of $\lambda$.
The enhanced population of
low distance states of gluons is due to Bose statistics and the color-Coulomb attraction.
In contrast, the depletion of the small
distance range for quarks is a consequence of the Pauli principle.
In an ideal Fermi gas $g(r)$ equals zero for particles with the
same spin projections and colors, while for particles with different colors and/or
opposite spins the PDF equals unity in the limit $r\to 0$. As a consequence,
the spin and color averaged PDF approaches $0.5$.
Such a low-distance behavior is also observed in a nonideal dense
astrophysical electron-ion plasma and in a nonideal  electron-hole plasmas in semiconductors \cite{bonitz_jpa06,filinov_pre07}.
The depletion of the probability of quasiparticles at small distances results in its enhancement at intermediate
 distances. This is the reason for the corresponding PDF maxima. 

At small distances, $r\le 0.3\sigma$, a strong increase is observed
in all PDF's of particles of different type (top right panel of Fig. \ref{fig:PDFC}),
which resembles the behavior of the gluon-gluon PDF.
This increase
is a clear manifestation of an effective pair attraction 
of quarks and antiquarks as well as quarks (antiquarks) and gluons.
This attraction suggests that the color vectors of nearest 
neighbors
of any type are anti-parallel.
%
%
If this explanation is correct can be verified by means of
 {\em color pair distribution functions} (CPDF) defined as 
\begin{eqnarray}\label{c-def}
c_{ab}(R_1-R_2)=
\frac{1}{Z N_q!N_{ \bar{q}}!N_g!}
\sum_{\sigma}\int
dr dQ\,\langle Q^a_1|Q^b_2 \rangle \delta(R_1-r^a_1)\delta(R_2-r^b_2)\rho(r,Q, \sigma ;\beta),
\end{eqnarray}
%
which are shown in the lower panels of Fig.  \ref{fig:PDFC}.
All CPDF's 
turn out to be negative at small distances, indicating anti-parallel orientation of the color vectors
of neighboring quasiparticles.
The minimum of $c_{qq}$ close to $r=0.2\sigma$ corresponds to the maximum observed in $g_{qq}$.
The deep minimum in the
gluon CPDF
at small distances results from the Bose statistics and 
complies with the high maximum of the gluon PDF $g_{gg}$.
%
\begin{figure}[htb]
\vspace{0cm} \hspace{0.0cm}
\includegraphics[width=6.8cm,clip=true]{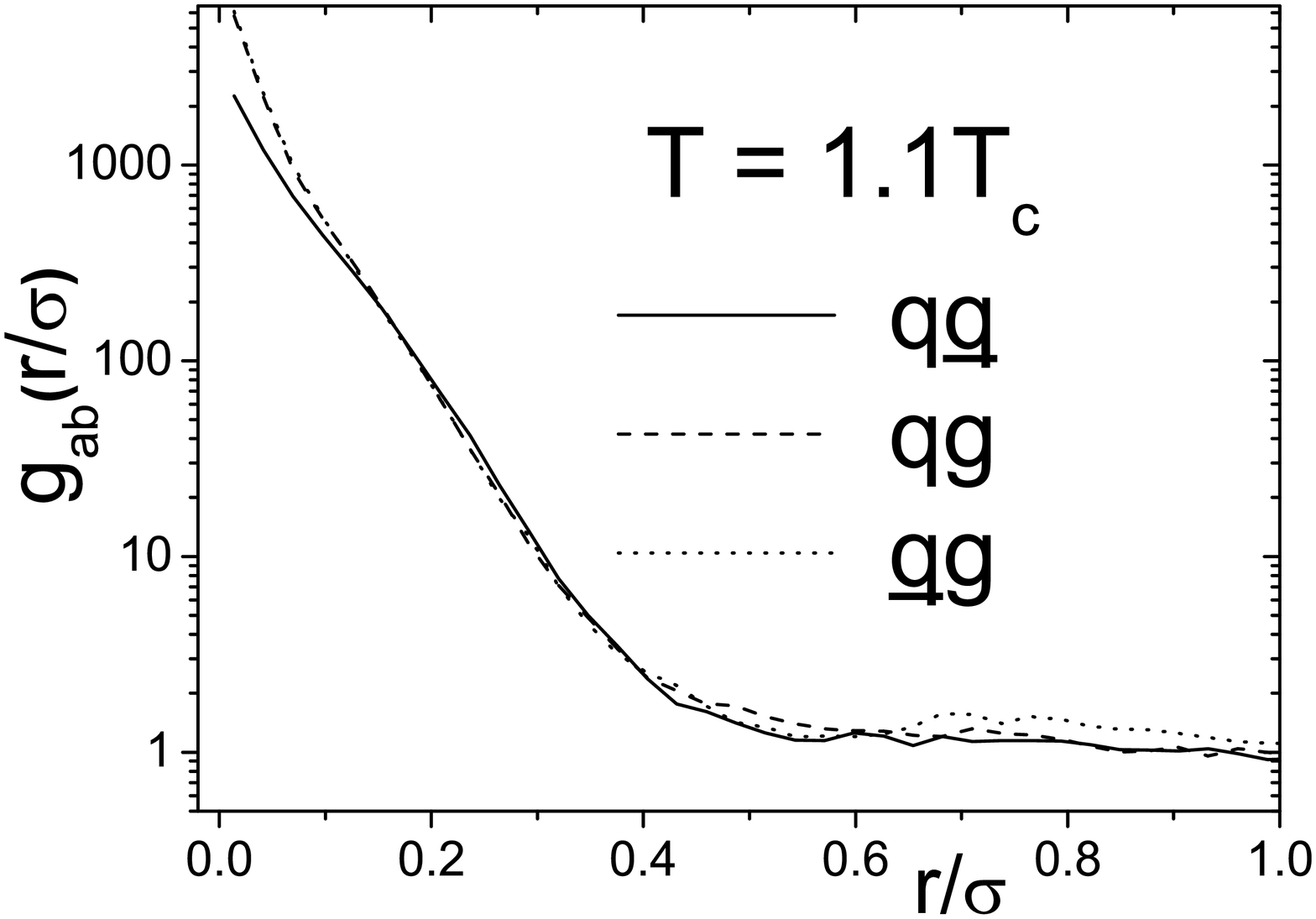}
\includegraphics[width=6.8cm,clip=true]{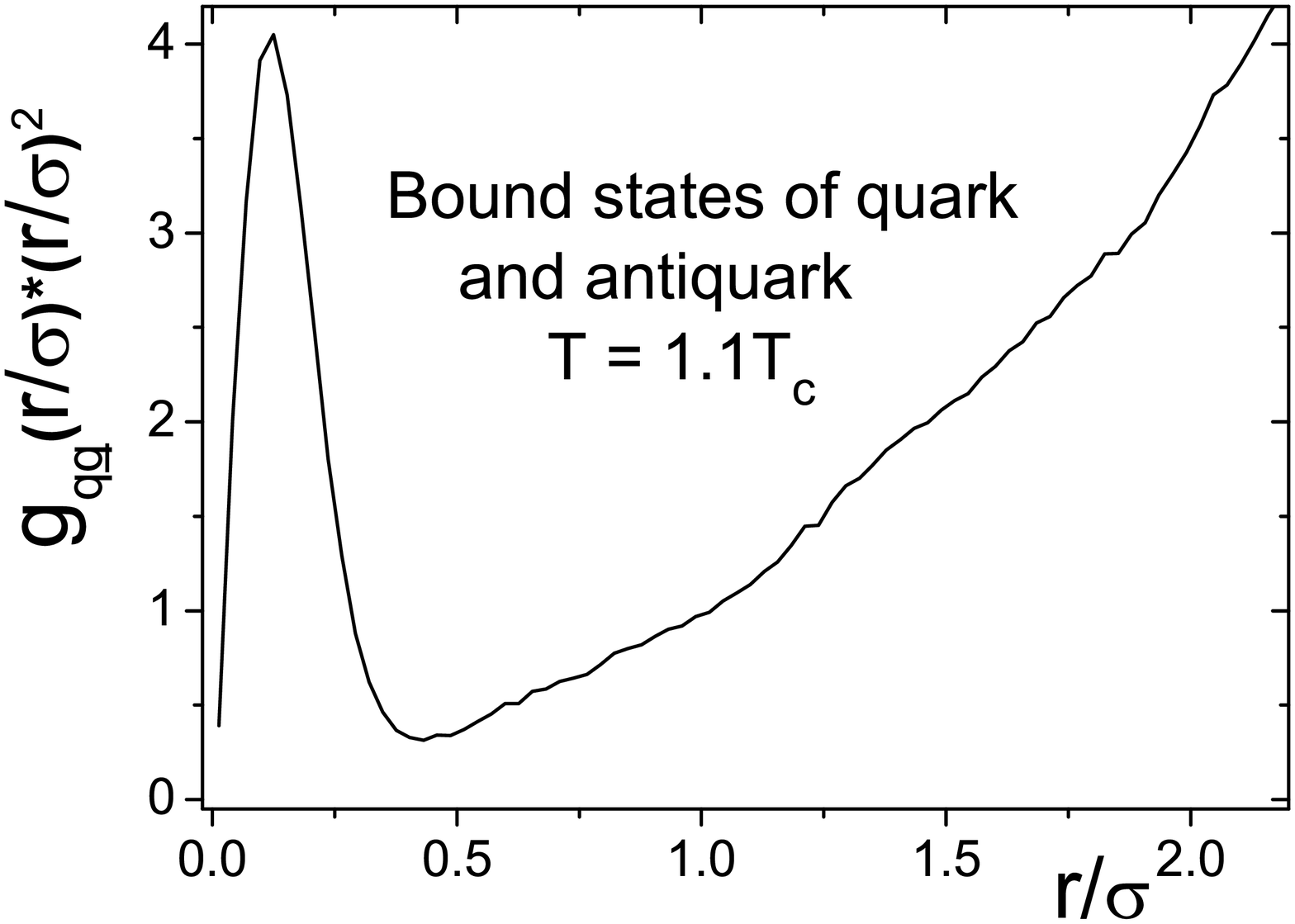}
\includegraphics[width=6.8cm,clip=true]{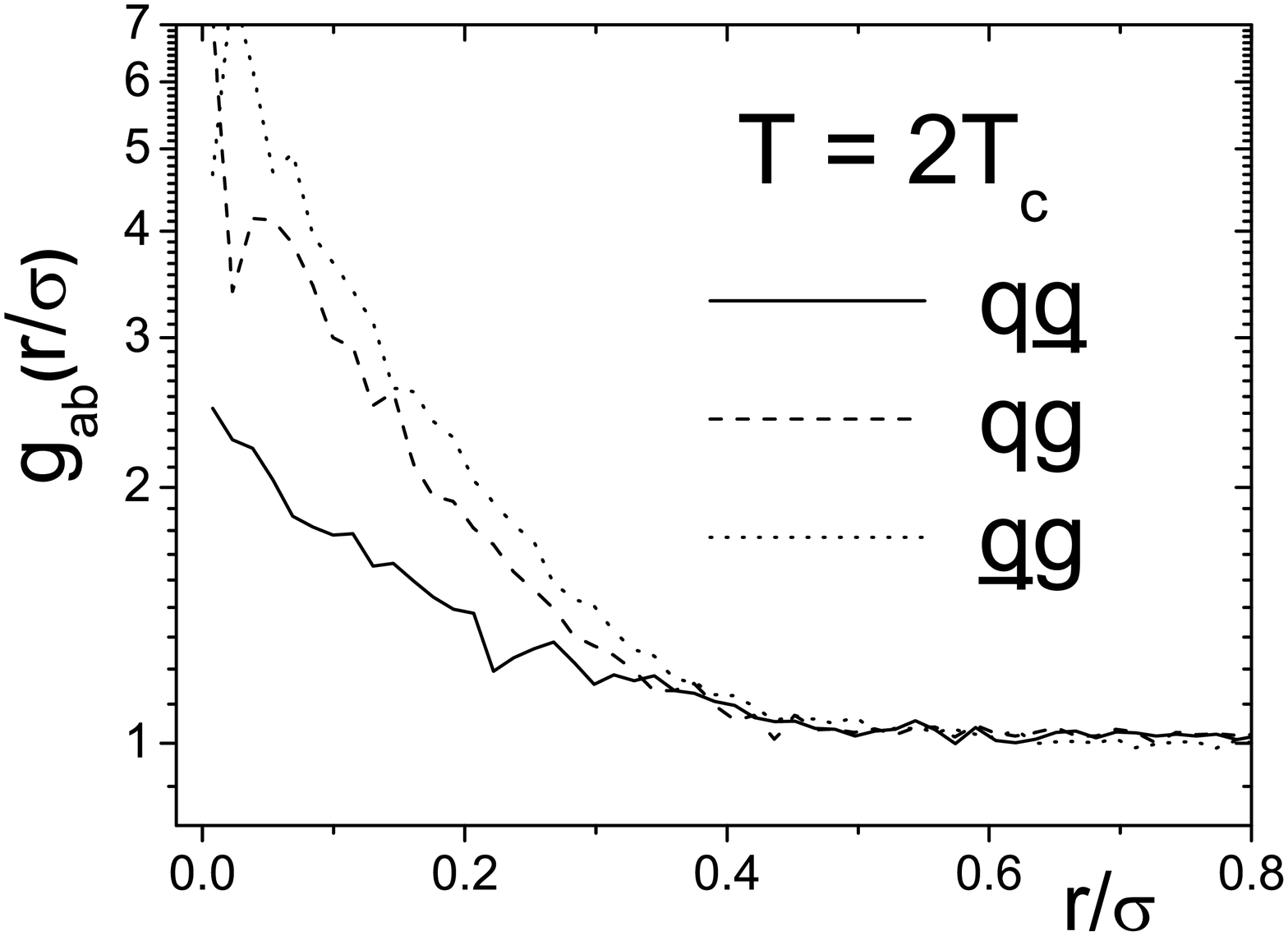}
\includegraphics[width=6.8cm,clip=true]{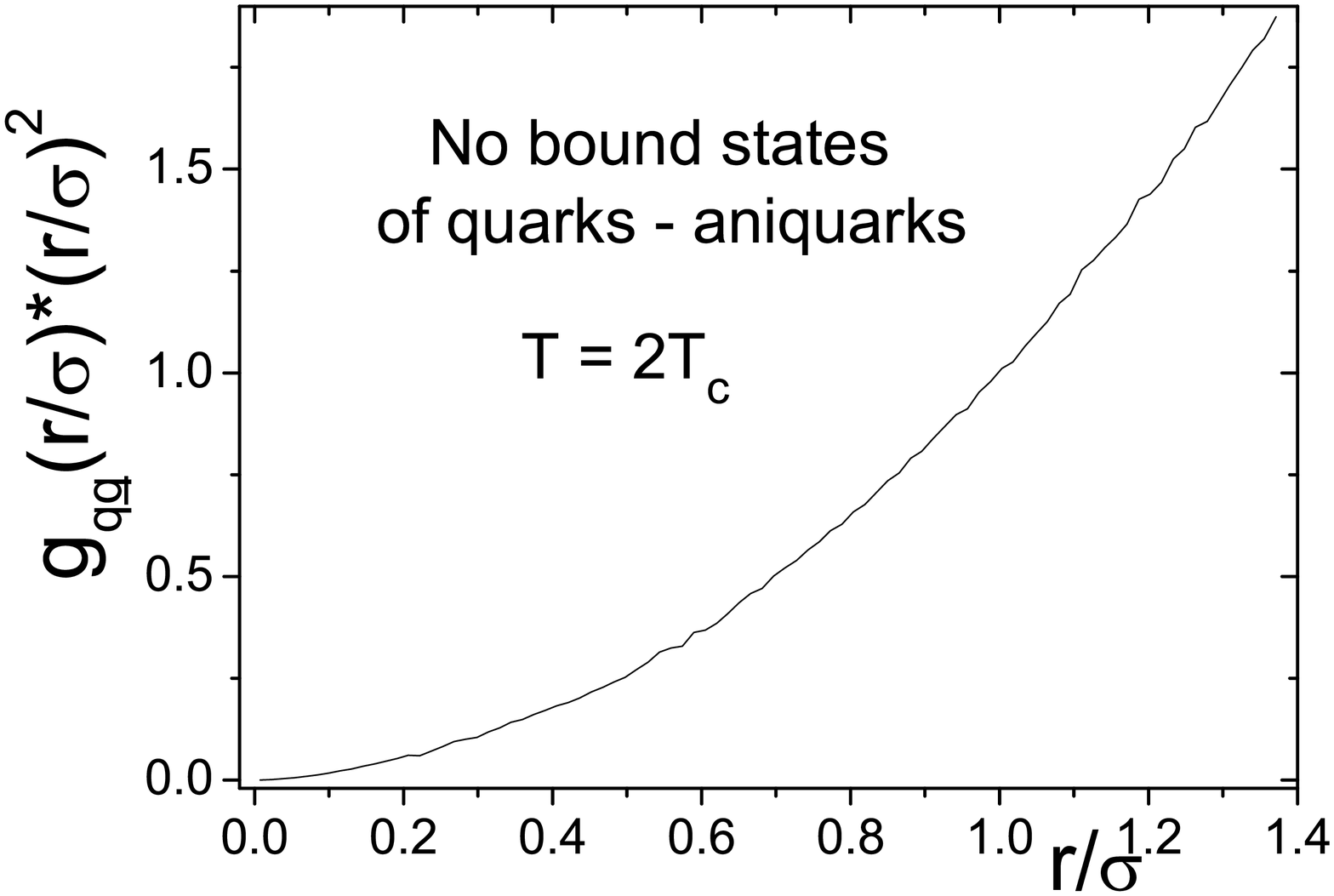}
\caption{
Pair distribution functions at two different temperatures $T$
(left) and quark-antiquark PDF multiplied by
distance squared (right).
}
\label{fig:PDBLSC}
\end{figure}
%

Thus, at $T/T_c=3$
we observe signs of a spatial ordering,
cf. the peak of the quark PDF around $r/\sigma=0.1-0.2$,
which may be interpreted as emergence of liquid-like behavior of the QGP.
The QGP lowers its total energy by minimizing the color
Coulomb interaction energy via a spontaneous ``anti-ferromagnetic'' ordering of color vectors. 
This gives rise to a clustering
of quarks, antiquarks and gluons.

Fig. \ref{fig:PDBLSC} presents PDF's of the identical particles
for two temperatures $T=1.1T_c$ and $T=2T_c$ (left panels).
These PDF's can be formed either by correlated scattering states  or by bound states of quasiparticles,
depending on the relative fractions of these states.
In a strict sense, however,
there is no clear subdivision into bound and free ``components''
due to the mutual overlap of the quasiparticle clouds. In addition, there exists no
rigorous criterion for a bound state at high densities due to the strong effect of the surrounding
plasma. 
Nevertheless, a rough estimate
of the fraction of quasiparticle bound states can be obtained
by  the following reasonings. The product $r^2 g_{ab}(r)$
has the meaning of a probability to find
a pair of quasipartices at a distance $r$ from each other.
On the other hand, the corresponding quantum mechanical
probability is the product of $r^2$ and the two-particle Slater sum
\begin{equation}
\label{slsm}
\Sigma_{ab}=8\pi
^{3/2}\lambda_{ab}^{3}\sum_{\alpha}|\Psi_{\alpha}(r)|^{2}\exp(-\beta
E_{\alpha})
 = \Sigma_{ab}^{d}+\Sigma_{ab}^{c},
\end{equation}
where $E_{\alpha}$ and $\Psi_{\alpha}(r)$ are
the energy (without center of mass energy) and the wave 
function of a quasiparticle pair, respectively, and 
$\lambda_{ab}=\sqrt{2\pi\hbar^2 \beta (m_{a}+m_{b})/(m_{a}m_{b})}$.
$\Sigma_{ab}$ is, in essence,
the diagonal part of the corresponding density matrix.
The summation runs over all 
states $\alpha$ of
the discrete ($\Sigma_{ab}^{d}$) and
continuous ($\Sigma_{ab}^{c}$)
spectrum. 

At temperatures smaller than the binding energy and
distances smaller than or of the order of several bound state radii, the main
contribution to the Slater  sum comes from bound
states.
In the electromagnetic plasma it was found that
the product
$r^2\Sigma_{ab}^d$ is sharply peaked at distances around the Bohr radius in this case.
Similarly, at low temperature, $r^2g_{q\bar{q}}(r)$ forms a pronounced maximum near
$r=0.2$ fm which can be interpreted as the radius of a bound $q\bar{q}$ pair.
Thus, our calculations support the existence of bound states of medium-modified 
(massive) quarks and gluons at moderate temperatures, i.e. just above $T_c$,
proposed in Ref. \cite{Yukalov97} and later in Refs. \cite{Shuryak03,Brown05}
based on results from lattice QCD calculations
of spectral functions \cite{Asakawa01,Karsch03}. 
%
With the temperature rise these bound states dissolve much faster than it was assumed in
\cite{Shuryak03,Brown05}, which complies with the analysis of Ref. \cite{Koch05}. 
Indeed, at the temperature of $T= 2 T_c$ the bound states completely disappear
(see right lower panel of Fig. \ref{fig:PDBLSC}).

Interesting observations can be done from the analysis of
the potential of average force (PAF) defined as the logarithm of  the related PDF,
$U_{ab}(r,T) = -T \ln g_{ab} (r,T)$.
This definition is motivated by the PDF virial expansion in terms of bare potential (like color Kelbg potential).
Near the QGP phase transition the PAF 
is a linear function at distances smaller than the bound state radius.
This suggests that the bound states are bound by a string-like forces.
At larger distances the PAF 
can be very well approximated
by an exponentially screened Coulomb potential (Yukawa-type potential)
like that in the electromagnetic plasma.
\section{Conclusion}\label{s:discussion}

Quantum Monte Carlo simulations 
based on the quasiparticle picture of the QGP
are able to reproduce the lattice equation of state (even near the critical temperature) and also
yield valuable insight into the internal structure of the QGP.
Our results indicate that the QGP reveals liquid-like (rather than gas-like) properties even at the
highest considered temperature of $3T_c$. At temperatures just above $T_c$ we have found that
bound quark-antiquark states still survive. These states are bound by effective string-like forces.
Quantum effects turned out to be of prime importance in these  simulations.



Our analysis is still too simplified and incomplete. It is still confined only to the case of zero
baryon chemical potential. The input of the model also requires refinement.
Work on these problems is in progress. We have also performed first simulations of dynamic properties of
the QGP  based on quantum Wigner dynamics. In particular, these allow us to deduce the viscosity of
the QGP. However, the brief format of the present contribution does not allow us to report on the
respective results.

%

We acknowledge stimulating discussions with D.~Blaschke, M.~Bleicher, R. Bock, B.~Friman,
C. Ewerz, D.~Rischke, and H.~Stoecker.
Y.I. was partially supported by the Deutsche  Forschungsgemeinschaft
(DFG projects 436 RUS 113/558/0-3 and WA 431/8-1),
the RFBR grant 09-02-91331 NNIO\_a, and grant NS-7235.2010.2.


\bibliographystyle {apsrev}

\begin{thebibliography}{90}
\bibitem{shuryak08}
E. Shuryak, Prog. Part. Nucl. Phys. {\bf 62}, 48 (2009).
\bibitem{Lattice09}
A. Bazavov, et al., 	arXiv:0903.4379 [hep-lat].
\bibitem{Fodor09}
Z. Fodor and S.D. Katz, arXiv:0908.3341 [hep-ph].
\bibitem{Csikor:2004ik}
  F.~Csikor, {\em et al.}, 
  JHEP {\bf 0405}, 046 (2004).
\bibitem{LM105}
D.F.~Litim and C.~Manuel,
Phys. Rev. Lett. {\bf 82}, 4981 (1999); 
Nucl.Phys. {\bf B 562}, 237 (1999); 
Phys. Rev. {\bf D 61}, 125004 (2000); 
Phys. Rep. {\bf 364}, 451-539 (2002).
\bibitem{Bleicher99}
M. Hofmann, {\em et al.}, 
 Phys. Lett. {\bf B 478}, 161 (2000). 
%
\bibitem{shuryak1} B.A. Gelman, E.V. Shuryak, and I. Zahed, Phys. Rev. C \textbf{74}, 044908 (2006);
ibid. \textbf{74}, 044909 (2006).
%
\bibitem{Zahed}
S. Cho and I. Zahed,
Phys. Rev. {\bf C 79}, 044911 (2009); 
 Phys. Rev. {\bf C80}, 014906 (2009); 
arXiv:0910.2666 [nucl-th];
arXiv:0910.1548 [nucl-th];
arXiv:0909.4725 [nucl-th];
K. Dusling and I. Zahed,  Nucl. Phys. {\bf A 833}, 172 (2010). 
%
%
%
\bibitem{thoma04} M.H.~Thoma, IEEE Trans. Plasma Science {\bf 32}, 738 (2004)
%
%

%
%
\bibitem{kelbg} G.~Kelbg, Ann. Physik (Leipzig) {\bf 12}, 219 (1962); {\bf 13}, 354 (1963).
%
\bibitem{dusling09} The idea to use a Kelbg-type effective potential also for quark matter
was independently proposed  by K.~Dusling
and C.~Young, arXiv:0707.2068v2. However, their potentials are limited to weakly nonideal systems.
%
\bibitem{bonitz_jpa03} M. Bonitz {\em et al.},
J. Phys. A 
{\bf 36}, 5921 (2003);
Phys. {\bf 15}, 055704 (2008).
%
%
\bibitem{filinov_ppcf01} V.S. Filinov, M. Bonitz, W. Ebeling, and V.E. Fortov,
 Plasma Phys. Control. Fusion {\bf 43}, 743 (2001).
%
\bibitem{filinov_jetpl00} V.S. Filinov, M. Bonitz, and V.E. Fortov,
 JETP Lett. {\bf 72}, 245 (2000). 
%
\bibitem{bonitz_prl05} M. Bonitz, {\em et al.}, 
Phys. Rev. Lett. {\bf 95}, 235006  (2005).
%
\bibitem{filinov_jpa03} V.S. Filinov, {\em et al.}, 
J. Phys. A: Math. Gen. {\bf 36}, 6069 (2003).
%
\bibitem{bonitz_jpa06} M. Bonitz, {\em et al.}, 
J. Phys. A: Math. Gen. {\bf 39}, 4717 (2006).
%
\bibitem{filinov_pre07} V.S.~Filinov, {\em et al.}, 
 Phys. Rev. E {\bf 75} , 036401 (2007).
%
\bibitem{Filinov:2009pimc}
V.S. Filinov, {\em et al.}, 
Contrib. Plasma Phys.,  \textbf{49}, 536 (2009).
%
%
\bibitem{Lattice02}
P. Petreczky, {\em et al.}, 
Nucl. Phys. Proc. Suppl. {\bf 106}, 513 (2002).
%
\bibitem{LiaoShuryak} J. Liao and E.V. Shuryak, Phys. Rev. D \textbf{73}, 014509 (2006).
%
\bibitem{Wong} S.K. Wong, Nuovo Cimento \textbf{A 65}, 689 (1970).
%
\bibitem{Rich} J.L. Richardson, Phys. Let. \textbf{82 B}, 272 (1979).
\bibitem{feynm}  R.P.~Feynman, and A.R.~Hibbs, {\it Quantum
Mechanics and Path Integrals}, McGraw-Hill, New York, Moscow 1965.
%
\bibitem{zamalin} V.M.~Zamalin,  G.E.~Norman, and V.S.~Filinov,
{\em The Monte Carlo Method in Statistical Thermodynamics}, Nauka,
Moscow 1977 (in Russian).
%
%
\bibitem{afilinov_pre04} A. Filinov, {\em et al.}, 
Phys. Rev. E {\bf 70}, 046411 (2004)
\bibitem{Peshier96}
A. Peshier, B. Kampfer, and O.P. Pavlenko, Phys. Rev. D {\bf 54}, 2399 (1996).
%
\bibitem{Ivanov05}
Yu.B. Ivanov, V.V. Skokov, and V.D. Toneev, Phys. Rev. D {\bf 71}, 014005 (2005).
%
\bibitem{Karsch09a}
F. Karsch and M. Kitazawa, arXiv:0906.3941v1 [hep-lat].
%
\bibitem{Prosperi}
G. M. Prosperi, M. Raciti, and C. Simolo,
Prog. Part. Nucl. Phys. {\bf 58}, 387 (2007). 

\bibitem{Karsch04}
O. Kaczmarek, {\em et al.}, 
Phys. Rev. D {\bf 70}, 074505 (2004); Erratum-ibid. D {\bf 72}, 059903 (2005).

\bibitem{rinton} A.V.~Filinov, V.S.~Filinov, Yu.E.~Lozovik and M.~Bonitz,
in: ``Introduction to Computational Methods for Many-Body Physics'',
M. Bonitz and D. Semkat (eds.), Rinton Press, Princeton (2006).
%
\bibitem{Yukalov97}
V. I. Yukalov and E. P. Yukalova,
Physica A {\bf 243}, 382 (1997);
Fiz. Elem. Chastits At. Yadra {\bf 28}, 89 (1997).
%
\bibitem{Shuryak03} E.V. Shuryak and I. Zahed, Phys. Rev. C {\bf 70}, 021901 (2004);
Phys. Rev. D {\bf 70}, 054507 (2004).
\bibitem{Brown05} G.E. Brown, B.A. Gelman, and M. Rho, arXiv:nucl-th/0505037.
%
\bibitem{Asakawa01}  M. Asakawa, T. Hatsuda and Y. Nakahara, Prog. Part.
Nucl. Phys. {\bf 46}, 459 (2001); 
Nucl. Phys.
{\bf A 715}, 863 (2003); 
M. Asakawa and T. Hatsuda, Phys. Rev. Lett. {\bf 92}, 012001 (2004). 
\bibitem{Karsch03}  S. Datta, {\em et al.}, 
Rev. D {\bf 69}, 094507 (2004). 
%
\bibitem{Koch05}
V. Koch, A. Majumder, and J. Randrup,  Phys. Rev. Lett. {\bf 95}, 182301 (2005).
%
%
\end{thebibliography}

\end{document}